Title: Active Damping of Power Oscillations Following Frequency Changes in Low Inertia Power Systems

Authors: Marios Zarifakis, William T. Coffey, Yuri P. Kalmykov, Serguey V. Titov, Declan Byrne, Stephen J. Carrig

Published in: IEEE Transactions on Power Systems (Early Access)

URL: https://ieeexplore.ieee.org/abstract/document/8693580

Date of Publication: 17 April 2019

Print ISSN: 0885-8950

Electronic ISSN: 1558-0679

DOI: 10.1109/TPWRS.2019.2911845

Publisher: IEEE

Funding Agency: Electricity Supply Board; Science Foundation Ireland

# Active Damping of Power Oscillations Following Frequency Changes in Low Inertia Power Systems

Marios Zarifakis, William T. Coffey, Yuri P. Kalmykov, Serguey V. Titov, Declan J. Byrne and Stephen J. Carrig

*Abstract*— The absolute requirement to increase the amount of energy generation from renewable sources e.g. predominantly asynchronously connected wind turbines and photovoltaic installations, may in practice during transient events (where frequency changes are examined) excite oscillatory response of the power output of large grid connected synchronous-generators. The response of such generators must be controlled either by varying the applied torque of a turbine or by altering the electromagnetic torque in the airgap. Choosing the latter, the adequacy of a voltage regulator, particularly that of the embedded *Power System Stabilizer (PSS)* circuit, is investigated using the IEEE PSS1A model for the automatic voltage regulator of a synchronous generator driven by a gas turbine. The response is obtained via closed form analytic solutions for both small (linear) and large (nonlinear) scale transient events in the energy grid system. In tandem with the analytical study, the behavior simulated with a computer model from *MatLab-SimPowerSystems* is reviewed.

*Index Terms*—Control system synthesis, Power generation control, Power system protection, Power system stability, Power system transients, Rate of change of frequency or ROCOF, Renewable energy sources, Synchronous generators.

## I. Nomenclature

| | |
|---|---|
| $a_0, b_0, V_\infty$ | Constants describing $V_{in}$ for a linear response. |
| $a_1, b_1, c_1$, etc. | Coefficients for intermediate and output signals. |
| $a_{0j}$ | Weighting for eigenfunctions. |
| $\beta_{grid}, \beta_{gen}$ | Reduced damping coefficient. |
| $\beta$ | Damping coefficient for the rotor angle equation. |
| $\mathbf{C}(t)$ | Vector describing current state of the system. |
| $\delta(t)$ | Generator rotor angle. |
| $\delta_\mathrm{I}, \delta_\mathrm{II}$ | Initial and final rotor angles. |
| $f(t)$ | Frequency of the bus voltage. |
| $J_{gen}, J_{grid}$ | Rotational inertia. |
| $K_S, K_{PR}, K_{PS}$ | Gain parameter for the PSS1A. |
| $K_D$ | Damping coefficients for the cage model. |
| $K_{gen}^K, K_{grid}^K$ | Damping coefficients for the Kuramoto-like model. |
| $\lambda$ | Decay rate for the linear response. |
| $\lambda_j$ | Eigenvalue for nonlinear response. |
| $\Omega$ | Unperturbed rotor angular speed. |
| $\omega_0$ | Frequency of oscillations for the linear response. |
| $\omega_e$ | Sine wave envelope for oscillations, see (13). |
| $p$ | Number of field poles in the generator. |
| $P_{el}(t)$ | Electrical output power. |
| $P_{max}$ | Maximum electrical output power. |
| $T_1, T_2$, etc. | Time constants for the PSS1A and AVR. |
| $\tau_{grid}, \tau_{gen}$ | Applied torque. |
| $\bar{\tau}_{grid}, \bar{\tau}_{gen}$ | Reduced torque coefficient. |
| $\tau_r$ | Torque coefficient for the rotor angle equation. |
| $\tau_{el\max}$ | Maximum electromagnetic torque in the air gap. |
| $\theta_{gen}, \theta_{grid}$ | Angle of generator and grid respectively. |
| $V_{in}$ | Input signal to the PSS1A. |
| $V_{PSS}$ | Output signal of the PSS1A. |
| $V_{PSS}$ | Output signal of the AVR. |
| $V_1, V_2$, etc. | Intermediate signals for the PSS1A and AVR. |
| $x$ | Grid to generator inertia ratio. |
| $\mathbf{X}$ | System matrix describing the generator and grid. |
| $\xi$ | Maximum electromagnetic torque coefficient for the rotor angle equation. |
| $\xi_{grid}, \xi_{gen}$ | Reduced maximum electromagnetic torque coefficient. |
| $\xi_\mathrm{I}, \xi_\mathrm{II}$ | Initial and final maximum electromagnetic torques. |

Manuscript received xxx, 2018; revised xxx, 2019; accepted xxx, 2019. Date of publication xxx, 2019. (Corresponding author: Declan J. Byrne.)
S. V. Titov acknowledges the Electricity Supply Board for financial support.
D. J. Byrne acknowledges Science Foundation Ireland for financial support. This publication has emanated from research conducted with the financial support of Science Foundation Ireland under Grant number 17/IFB/5420.
M. Zarifakis and S. J. Carrig are with the Electricity Supply Board, Engineering and Major Projects, Dublin 3, Ireland (e-mail: marios.zarifakis@esb.ie, stephen.carrig@esb.ie).
W. T. Coffey and D. J. Byrne are with the Department of Electronic and Electrical Engineering, Trinity College, Dublin 2, Ireland (e-mail: wcoffey@mee.tcd.ie; byned12@tcd.ie).
Yu. P. Kalmykov is with the Laboratoire de Mathématiques et Physique (EA 4217), Université de Perpignan Via Domitia, F-66860, Perpignan, France (e-mail: kalmykov@univ-perp.fr).
S. V. Titov is with the Kotel'nikov Institute of Radio Engineering and Electronics of the Russian Academy of Sciences, Vvedenskii Square 1, Fryazino, Moscow Region, 141190, Russia (email: pashkin1212@yandex.ru).

## II. Introduction

THE ever-present requirement to decarbonize energy generation and therefore to increase energy levels from *Renewable Energy Sources* (RES) means that wind turbines and solar photovoltaic installations have become major energy pool contributors. Invariably studies of grids with high penetration of RES (particularly isolated island grids such as Ireland) indicate that the increase of these sources weakens the ability of the frequency in the transmission and distribution system to remain stable after transient disturbances [1]-[5]. The reason being that RES, contrary to conventional *synchronously* grid-connected turbo generators of gas, oil or even coal fired power stations, are *asynchronously* connected causing the grid rotational inertia (due to the stored kinetic energy of the generators on the grid) to become low [1]-[5]. Therefore, compared to the infinite inertia grid, the response of the low inertia grid to a disturbance becomes significantly more unstable and new effects must be accounted for such as generator-grid feedback and increased Rate of Change of Frequency (ROCOF) [6] leading to oscillations in the entire transmission system. Hence maximizing the RES level on a transmission system without compromising the safety and integrity of existing generator assets must be investigated.

Recent publications exist exploring the effect of increasing grid RES levels on its rotational inertia and stability following a disturbance. This has usually been achieved by finding relevant physical characteristics using simulations on various testbeds, e.g., [1]-[4] simulated the dynamic response, [2] analysed the eigenvalue sensitivity, [5] investigated the effects on the rate of change of rotor speed, and [3], (using a five-machine reduced model to represent The Western Electric Coordinating council transmission grid) investigated inter-area power-flow oscillations. Additionally, methods have been proposed for tuning system parameters to account for high RES penetration. For example, in [4] controllers for doubly fed induction generators for wind farms were designed so that instabilities resulting from a disturbance on the wind farm could be prevented. Yet another method is Koopman mode decomposition [7],[8] which is relevant to the current paper as the nonlinear dynamic response of the system is represented as a sum of eigenfunctions in both cases, although the methods for obtaining them differ significantly (as discussed below).

Here we study how one may more rapidly stabilize the generator load angle $\delta$ following a disturbance by introducing an active control loop, so reducing the power oscillations of a grid-connected synchronous generator, in practice achieved by adding within the generator voltage regulator an additional control loop called a Power System Stabilizer (PSS) changing the excitation current in the rotor of a synchronous generator. Thus, by altering the magnetic field created by the excitation current, the torque in the airgap of the generator is controlled [9]. Although in general power system stabilizers reduce undesirable transient torsional oscillations of large turbine generator shafts, recent measurements on generators connected to the isolated transmission system of the island of Ireland exhibit *increased* power oscillations [10]. This unwanted phenomenon must be immediately addressed from both experimental and theoretical points of view so as to understand the implications for both generation assets and the grid itself.

Historically the dynamical models used to describe [11] grid systems assume that a grid has *infinite* inertia. However, recent measurements cannot be explained by this *Ansatz* as *generator-*

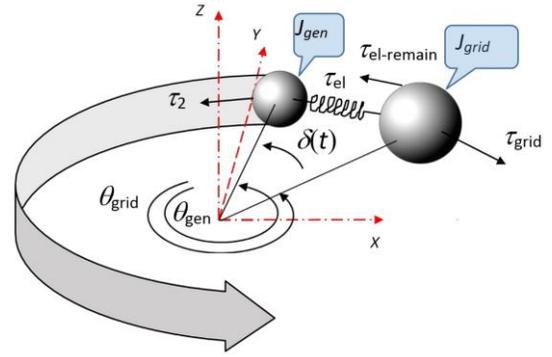

Fig. 1. Rotating torsional pendulum model

*grid feedback* is ignored. Therefore, we recently developed new dynamical methods based on a rotating double pendulum [10], [12]. Our model (v. Fig.1) [10], [12] is two rotating masses representing on the one hand the inertia $J_{gen}$ of the grid-connected synchronous generating unit and on the other the grid itself represented by the inertia $J_{grid}$. The work will be based on a recent paper [12] where appropriate dynamical equations for low inertia grids (as summarized in Appendix A) were written as differential-recurrence relations so that matrix algebra yields the relevant characteristics (based on methods developed in [13] and [14]). Specifically, the nonlinear response of the rotor angle to a disturbance is given as a sum of eigenfunctions which is our basis for critically examining the adequacy of the common power system stabilizers type PSS1A and its tuning.

One of the advantages of our method is that the results derived are analytic, yielding intuitive understanding of the effects of the nonlinear dynamics on the system. Additionally, we consider the response with respect to a single generator using a two-body system where the rest of grid acts as a single unit. Our solution is based on the equations of motion for a low inertia grid. Thus, we are not confined by system parameters. Additionally, as the solution completely describes the nonlinear dynamics, we can consider any size of fault. Furthermore, our method does not require a simulation to be run. For these reasons we believe that this method will be useful to practical engineers in the area of energy generation seeking to analyse the effects of PSS and AVR in low inertia grids.

The paper is arranged as follows. Firstly, a model is created in *MatLab* using transfer function blocks from the *Simulink* library, via the appropriate *Simulink embedded linear analysis tool* so yielding a Bode plot. Using *Simulink,* the generator and the corresponding overall transfer function are modelled, ultimately yielding a comprehensive model of the *entire circuit* valid for all parameter values. Next analytic solutions are obtained via *s*-plane analysis of the relevant cascaded block diagram of the PSS1A and the automatic voltage regulator (AVR) for both linear and nonlinear responses.

### III. MODELS FOR PSS BEHAVIOR

Our starting point is the block diagram of the Power System Stabilizer, Fig. 2 (a) [15], where the corresponding cascaded transfer functions of each block used for the analytic calculation

TABLE I

PARAMETERS FOR PSS1A

| Parameter | Value |
|---|---|
| $T_1$ | 0.4s |
| $T_2$ | 1.0s |
| $T_3$ | 0.1s |
| $T_4$ | 0.05s |
| $T_5$ | 2.0s |
| $T_6$ | 0.028s |
| $K_S$ | 0.8 |

of the response are shown explicitly. Commonly the input signals to the PSS1A will use output characteristics of the generator including the rotor speed deviation, the frequency deviation of the bus voltage or the electrical power output [15]. Following a disturbance, the oscillating component of these characteristics then supplies the input signal $V_{in}$ to the PSS1A circuit as shown in Fig. 2 (a).

In summary, the PSS model provides an input ($V_{PSS}$) to the AVR, ideally inducing *active* damping of the power oscillations due to the load angle oscillations in the airgap of the synchronous generator. A variety of stabilizing signals $V_{PSS}$ may be used depending on the particular design. However, we deliberately chose the signal generated by the PSS1A model with block diagram as in Fig. 2 (a). Here $V_{in}$ is the *input signal*, whereas $V_1$, $V_2$, and $V_3$ at each cascaded stage are called *intermediate signals*, $K_S$ is a factor of proportionality and $T_i$ are time constants as shown in Table 1 ultimately yielding the Bode plot of the transfer function as shown in Fig. 3 (a) via *MatLab Simulink*. The output of the PSS provides an input to the AVR (v. Fig. 2 (b)) and referring to the first block $T_N = 2$ s is the integration time of the regulator, $T_S = 1.8$ ms is the time constant of the bridge, $K_{PR}$ and $K_{PS}$ are constants of proportionality. For simplicity $K_{PR} = K_{PS} = 1$. Using *Simulink* to draw the Bode plot of the entire circuit including the PSS and AVR, we have Fig. 3(c). An image of this circuit is given in the online supplementary material.

## IV. ANALYTIC METHOD FOR A LINEAR TRANSIENT RESPONSE

Following [10] and [12] (see Appendix A), the dynamics of a generator connected to a low inertia power system can be described (using either a cage or Kuramoto-like model) via the equation of motion of the rotor angle $\delta(t)$, (A4) from Appendix A. For a two-pole synchronous generator the terms load angle and rotor angle are interchangeable. To study the cascaded transfer function analytically we select an input signal corresponding in general to an actual signal of the generator during a transient event capable of yielding the stabilizing signal $V_{PSS}$ in closed form. To analyze the linear transient response due to a sudden *small* change in the maximum electromagnetic torque $\tau_{el\max}$ [12] (see Appendix A) at the instant $t = 0$, this input is represented as the damped oscillation,

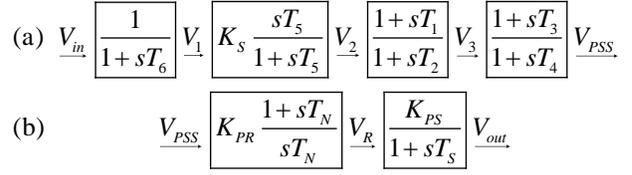

Fig. 2. Block diagram of (a) PSS1A model and (b) AVR.

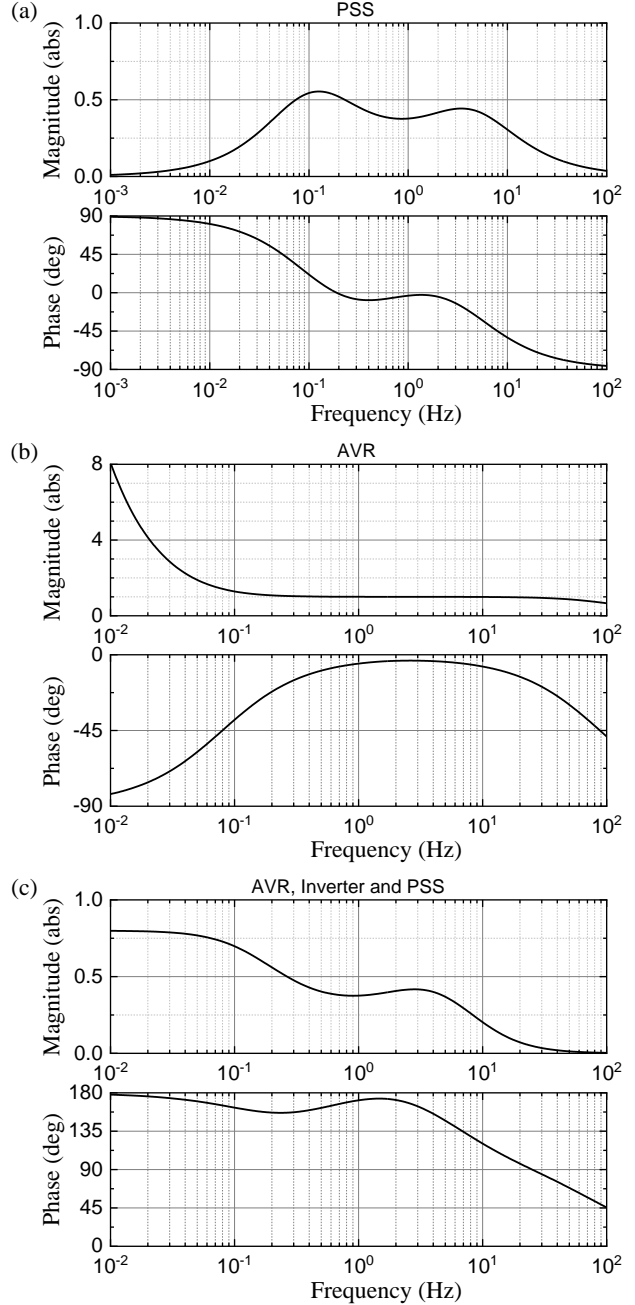

Fig. 3. Bode diagram of the transfer function: (a) PSS1A, (b) AVR, and (c) PSS1A and AVR.

$$V_{in} = a_0 e^{-\lambda t} \sin \omega_0 t + b_0 e^{-\lambda t} \cos \omega_0 t + V_\infty, \qquad (1)$$

where $V_\infty = V_{in}(t \to \infty)$ and the decay and frequency of the oscillations are [16]

$$\lambda = \beta/2, \quad \omega_0 = \sqrt{\xi \cos \delta_{II} - \beta^2/4}. \qquad (2)$$

The constants $a_0$, $b_0$, and $V_\infty$ can then be found explicitly as follows. The input to the PSS1A is usually the rotor angular speed deviation [15]. Therefore, since the rotor angular speed is given by (A6), the deviation (for an infinite inertia grid) $-\dot\delta(t)$ is simply given by (1) with

$$a_0 = \omega_0^{-1}\xi(\delta_\mathrm{I} - \delta_\mathrm{II})\cos\delta_\mathrm{II},\ b_0 = 0,\ V_\infty = 0. \quad (3)$$

Another typical input to the PSS1A is the frequency deviation of the bus voltage [15] which, using (1) and (3), is [9]

$$f(t) = p\omega(t)/2,\ \Delta f(t) = -p\dot\delta(t)/2, \quad (4)$$

where $p$ is the number of field poles in the generator. Yet another possible input is the electrical power output [15] written as [9]

$$P_{el}(t) = P_{\max}\sin\delta(t). \quad (5)$$

For disturbances with small rotor angle deviation $P_{el}(t)$ has the form of (1), viz.,

$$\begin{aligned}P_{el}(t) &= P_{\max}\sin\delta_\mathrm{II} + P_{\max}(\delta_\mathrm{I} - \delta_\mathrm{II})\cos\delta_\mathrm{II}\\ &\times e^{-\lambda t}\left[(\beta/2\omega_0)\sin\omega_0 t + \cos\omega_0 t\right].\end{aligned} \quad (6)$$

Thus since $V_{in}(t)$ is written in the generalized form of (1) it may now be used for various commonly used input signals. We consider the more relevant case of large rotor angle deviation (nonlinear transient response) in Section 4.

Now, each intermediate signal can also be represented (in linear transient response) as the sum of decaying terms (v. Appendix B), i.e., the output of the PSS1A is

$$\begin{aligned}V_{PSS}(t) &= (a_4\sin\omega_0 t + b_4\cos\omega_0 t)e^{-\lambda t} + c_4 e^{-t/T_6}\\ &+ d_4 e^{-t/T_5} + e_4 e^{-t/T_2} + f_4 e^{-t/T_4}\end{aligned} \quad (7)$$

where the coefficients are given in Appendix B. The output of the PSS is now used as an input to the AVR (v. the block diagram Fig. 2 (b)) so that the output of the AVR $V_{out}(t)$ is explicitly (v. Appendix B)

$$\begin{aligned}V_{out}(t) &= a_{out}e^{-\lambda t}\sin\omega_0 t + b_{out}e^{-\lambda t}\cos\omega_0 t + c_{out}e^{-t/T_6}\\ &+ d_{out}e^{-t/T_5} + e_{out}e^{-t/T_2} + f_{out}e^{-t/T_4} + g_{out}e^{-t/T_S} + s_R\end{aligned} \quad (8)$$

(for the various coefficients v. Appendix B).

Comparisons of $V_{in}$ and the signals $V_{PSS}$ and $V_{out}$ are shown in Fig. 4 in both the time and frequency domains with input signals corresponding to the rotor angular speed deviation and the electrical power output.

### V. ANALYTIC APPROACH TO THE NONLINEAR RESPONSE

Usually the input signal to the PSS1A cannot be represented as a single decaying oscillation. Therefore, we must also consider the nonlinear contributions during a transient event so that the input signal is then best described by the sum of eigenfunctions

$$V_{in} = \sum_j a_{0j}e^{\lambda_j t}. \quad (9)$$

Here $\lambda_j$ represent the eigenvalues of our dynamical system whereas the amplitudes $a_{0j}$ are determined from the corresponding eigenfunctions and the initial conditions [12]. As $V_{in} \in \mathbb{R}$, for any $\lambda_j, a_{0j} \in \mathbb{C}$ an equivalent conjugate term must exist, whereas for any $\lambda_j \in \mathbb{R}$, $a_{0j} \in \mathbb{R}$ also.

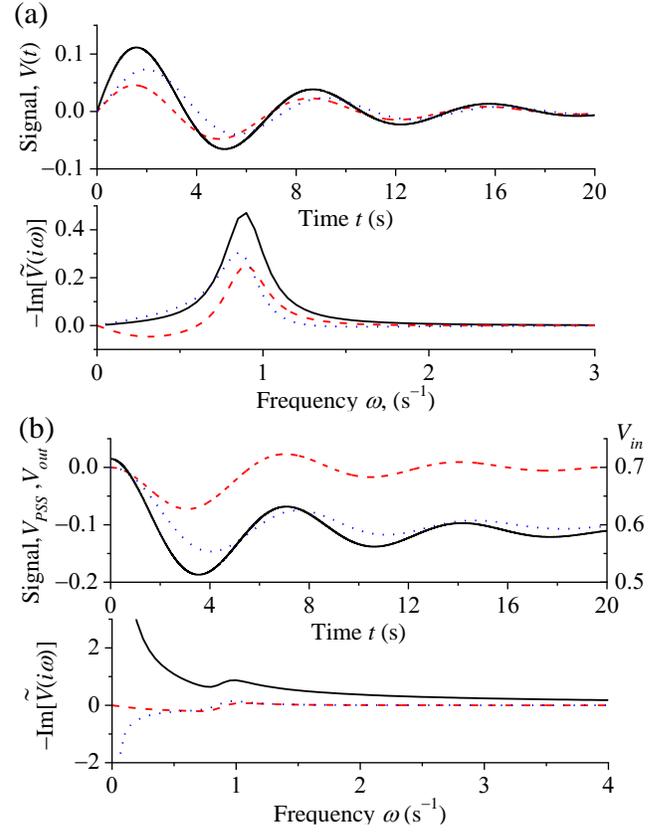

Fig. 4. (Color on line) $V_{in}(t)$ and imaginary part of its one-sided Fourier transforms(solid lines) corresponding to $-\dot\delta(t)$ (a) and $P_{el}(t)$ (b) for final coupling parameter $\xi_\mathrm{II} = 1$, initial angle $\delta_\mathrm{I} = \pi/4$, angular deviation $\Delta\delta = \pi/20$, and damping parameter $\beta = 0.3$. Dashed and dotted lines are, respectively, $V_{PSS}(t)$ and $V_{out}(t)$ and the corresponding imaginary parts of their one-sided Fourier transforms $\tilde V(s = i\omega)$.

An exact representation of the input signal for arbitrary system parameters and disturbance amplitude can be determined as in Ref. [12], where calculating the response of an energy generator within a low inertia grid (using either a cage or Kuramoto models [12]) following an abrupt change in the dynamical system parameters (e.g., tripping of generation plant) reduces to solving the first order matrix differential equation

$$\dot{\mathbf{C}}(t) + \mathbf{X}\mathbf{C}(t) = 0. \quad (10)$$

The system matrix $\mathbf{X}$ can then be used to determine the sum of eigenfunctions of (9). Thus we have *exact* equations in the form of (9) for the behavior of the PSS1A and AVR for *arbitrary* rotor angle disturbance amplitude and an arbitrary set of operating parameters for low inertia grid systems where we use either the Kuramoto or cage models for the response of the grid to a transient fault.

The calculations for the input signal given by (9) are as for (1) (v. Appendix B). Therefore, we shall not give them explicitly. The resultant signals $V_{PSS}(t)$ and $V_{out}(t)$ are

$$V_{PSS}(t) = \sum_j a_{4j}e^{\lambda_j t} + c_4 e^{-t/T_6} + d_4 e^{-t/T_5} + e_4 e^{-t/T_2} + f_4 e^{-t/T_4}, \quad (11)$$

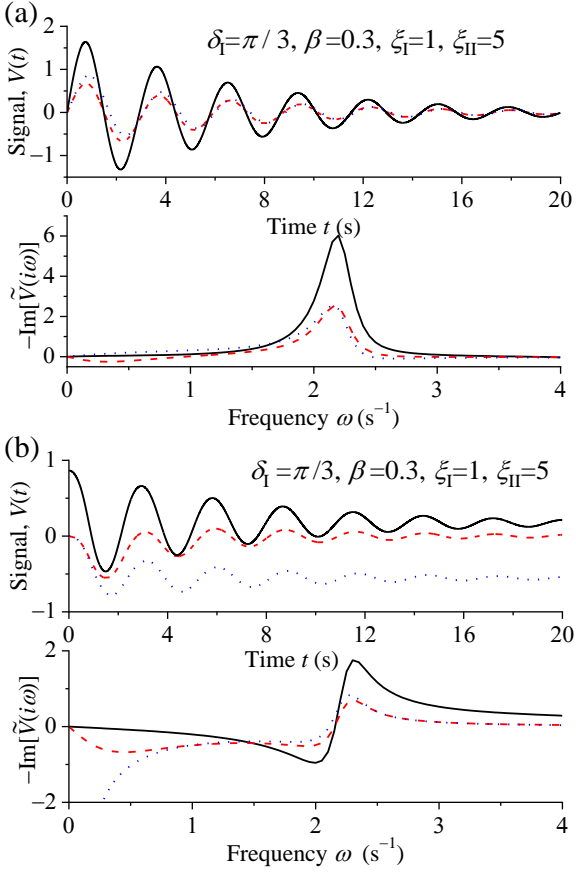

Fig. 5. (Color on line) Signals $V_{in}(t)$ (solid lines) corresponding to $-\dot{\delta}(t)$ (a) and . $P_{el}(t)$. (b) for initial coupling parameter $\xi_I = 1$, final coupling parameter $\xi_{II} = 5$, initial angle $\delta_I = \pi/3$, and damping parameter $\beta = 0.3$ (see Fig. 2 (c) of [12]). Dashed and dotted lines are, respectively, $V_{PSS}(t)$ and $V_{out}(t)$ and imaginary parts of their one-sided Fourier transforms.

$$V_{out}(t) = \sum_j a_{out,j} e^{\lambda_j t} + c_{out} e^{-t/T_6} + d_{out} e^{-t/T_5} \\ + e_{out} e^{-t/T_2} + f_{out} e^{-t/T_4} + g_{out} e^{-t/T_5} + s_R, \quad (12)$$

(for the coefficients, see Appendix C). Examples of $V_{in}(t)$, $V_{PSS}(t)$, and $V_{out}(t)$ are given in Fig. 5 in both the time and frequency domains with input signals corresponding to the rotor speed deviation and the electrical power output.

Fig. 5 shows the response of the PSS1A and AVR for an infinite grid inertia system. This instance, unlike finite grid inertia, is only of passing interest since renewable energy provided by non-synchronous generation sources, e.g., wind turbines and photovoltaic installations, cannot provide inertia to the grid as they are *asynchronously* connected to the power system. Referring now to finite inertia, the ever-present requirement to decarbonize energy generation means that this particular situation must be studied in detail. Therefore, the response of the PSS1A and AVR for various grid to generator inertia ratios, viz., $x = J_{grid}/J_{gen}$ (v. Appendix A) is needed. The results are shown for the Kuramoto-like model [12],[17]-[20] in Fig. 6(a) and for the cage model [12],[21]-[27] in Fig. 6 (b) (see Appendix A).

Our methods can also be applied to input signals other than decaying oscillations (generally corresponding to a single abrupt change in the dynamical system). Now, another input occurring in actual ROCOF events [10] comprises oscillations which initially increase and on attaining a peak amplitude then decrease. Analytically this response can be simulated by superimposing a series of square pulse waves on the applied torque [10]. In practical terms this response may occur if a sequence of abrupt changes to the dynamical system occurs in rapid succession (e.g., if a sudden change in system load leads to disconnection of generators, etc., creating a snowball effect).

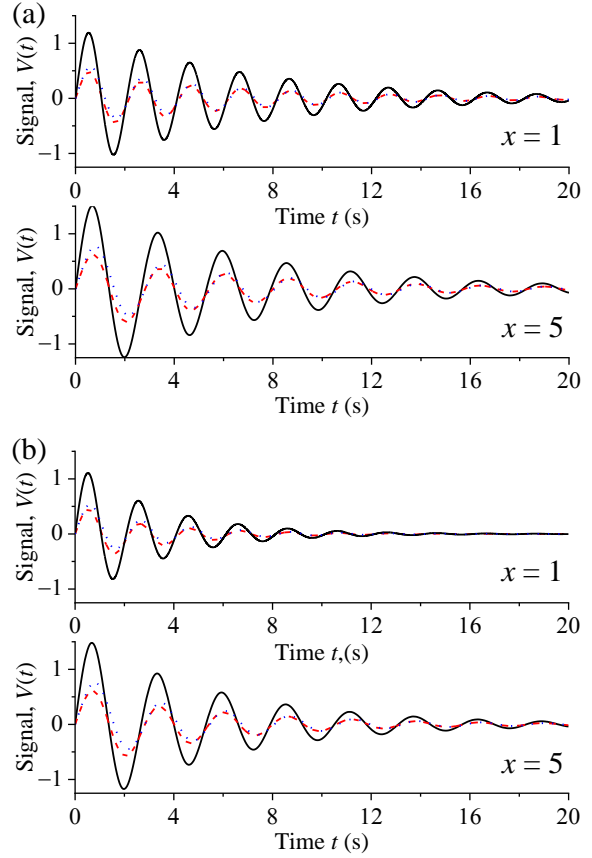

Fig. 6. (Color on line) Time dependence of $V_{in}(t)$ (solid lines) corresponding to $-(1+x^{-1})^{-1}\dot{\delta}(t)$, $V_{PSS}(t)$ (dashed lines), and $V_{out}(t)$ (dotted lines) for the Kuramoto-like (a) and the cage (b) models for various grid to generator inertia ratios $x = J_{grid}/J_{gen}$ and $\beta_2 = 0.3$, $\delta_I = \pi/3$, $\xi_I = 1$, $\xi_{II} = 5$.

We model such an input using a sine wave envelope

$$V_{in} = A\sin(\omega_e t)\sin(\omega_0 t) \quad (13)$$

existing only between $t = 0$ and $t = \pi/\omega_e$ so that in the $s$-domain

$$\tilde{V}_{in}(s) = A\big(2s\omega_e\omega_0(1+\cos(\pi\omega_0/\omega_e)e^{-s\pi/\omega_e}) \\ + \omega_e(s^2+\omega_e^2-\omega_0^2)\sin(\pi\omega_0/\omega_e)e^{-s\pi/\omega_e}\big) \quad (14) \\ \times \big((s^2+(\omega_0-\omega_e)^2)(s^2+(\omega_0+\omega_e)^2)\big)^{-1}.$$

Then we have as before *closed* form expressions for the stabilizing $V_{PSS}$ and output signals $V_{out}$, cf. the blocks of Fig. 2 which are used to determine the plots in Fig. 7.

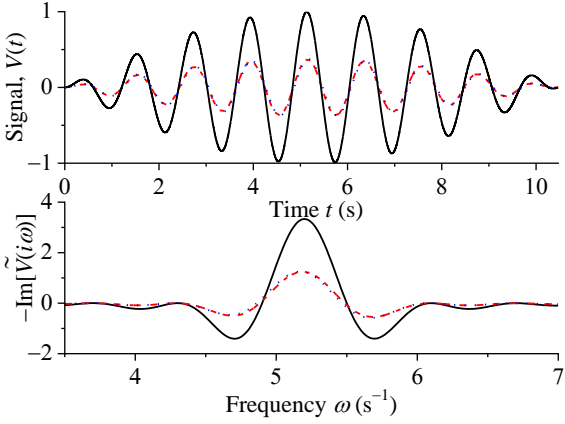

Fig. 7. (Color on line) Signals $V_{in}(t)$, Eq. (13) (solid line), $V_{PSS}(t)$ (dashed line), $V_{out}(t)$ (dotted line) and imaginary parts of their one-sided Fourier transforms for $A=1$, $\omega_e = 0.3$, $\omega_0 = 5.2$.

## VI. Conclusions

In conclusion, we have created an exact analytic solution describing the effect of the PSS1A and AVR in stabilizing the response of a generator to a transient fault based on our recently developed dynamical model of low inertia grids [10],[12]. Therefore, we will be able to explore the role of power system stabilizers on grids with high penetration of RES. Our method can consider the nonlinear response of the generator and grid to a large transient event and is not confined by the magnitude of the fault, the inertias of the grid and generator or the system parameters. Continuing this work, it is necessary to investigate whether or not the PSS1A values chosen in the PSS transfer functions can be adjusted so to ameliorate the response of the generator to ROCOF events *or* whether a different type of PSS, e.g., PSS2B or PSS4B, would yield even better results.

## Appendix

### A. Dynamics of a Grid-Connected Generator Following a Disturbance of the Low Inertia Grid

In [10] and [12], a generating station model based on a double torsional pendulum (called the cage model) is proposed, described by a system of coupled nonlinear differential equations and suitable for analysis of generator stability with either infinite or finite grid inertia. The cage model, where the grid has *finite* moment of inertia, has equations of motion for the generator and grid [10] (see Fig. 1)

$$J_{grid}\ddot{\theta}_{grid} + K_D(\dot{\theta}_{grid} - \dot{\theta}_{gen}) + \tau_{el\max}\sin(\theta_{grid} - \theta_{gen}) = \tau_{grid}, \quad (A1)$$

$$J_{gen}\ddot{\theta}_{gen} + K_D(\dot{\theta}_{gen} - \dot{\theta}_{grid}) + \tau_{el\max}\sin(\theta_{gen} - \theta_{grid}) = \tau_{gen}. \quad (A2)$$

Here $J_i$ denotes the relevant moment of inertia, $K_D$ is the damping coefficient, $\tau_{el\max}$ is the maximum electromagnetic torque in the air gap, $\tau_{gen}$ is the torque applied by the turbine to the generator, $\tau_{grid}$ is the resulting torque applied to the grid (sum of all turbine torques less the torques due to the loads and remaining generators $\tau_{el-remain}$ on the grid). Notice that the damping torques $\pm K_D(\dot{\theta}_{grid} - \dot{\theta}_{gen})$ in (A1) and (A2) exist only when the rotor angular velocity *differs* from the grid angular velocity. On introducing dimensionless parameters $\bar{\tau}_{grid} = \tau_{grid}/J_{grid}$, $\bar{\tau}_{gen} = \tau_{gen}/J_{gen}$, $\xi_{grid} = \tau_{el\max}/J_{grid}$, $\xi_{gen} = \tau_{el\max}/J_{gen}$, $\beta_{grid} = K_D/J_{grid}$, $\beta_{gen} = K_D/J_{gen}$, we rewrite the set of equations (A1) and (A2) as

$$\ddot{\theta}_i + \beta_i(\dot{\theta}_i - \dot{\theta}_j) + \xi_i\sin(\theta_i - \theta_j) = \bar{\tau}_i \ (i,j=grid,gen). \ (A3)$$

Subtracting the second equation ($i = gen$) of the set (A3) from the first one ($i = grid$) and introducing the rotor angle $\delta(t) = \theta_{grid}(t) - \theta_{gen}(t)$ yields

$$\ddot{\delta}(t) + \beta\dot{\delta}(t) + \xi\sin\delta(t) = \tau_r, \quad (A4)$$

where $\tau_r = \bar{\tau}_{grid} - \bar{\tau}_{gen}$, $\beta = \beta_{grid} + \beta_{gen}$, and $\xi = \xi_{grid} + \xi_{gen}$. This single-mass version of the model commonly used to analyze the dynamic response of a *synchronous* generator in an *infinite* grid [9]. The mechanical analog is a *driven damped pendulum*.

Next, to model the effects of *finite* grid inertia we introduce a new variable $x$, namely the ratio of the grid inertia to the generator inertia, $x = J_{grid}/J_{gen}$, allowing one to write the coupling and damping parameters as [12]

$$\xi = \frac{\tau_{el\max}}{J_{gen}}\left(\frac{x+1}{x}\right) \text{ and } \beta = \frac{K_D}{J_{gen}}\left(\frac{x+1}{x}\right). \quad (A5)$$

The rotor angular velocity is now given by [12]

$$\dot{\theta}_{gen}(t) = \Omega - \frac{x}{x+1}\dot{\delta}(t) \quad (A6)$$

where $\Omega = \dot{\theta}_{grid}(0) = \dot{\theta}_{gen}(0)$ corresponds to *unperturbed* (i.e., steady) rotation of grid and generator.

Equations (A1)-(A6) represent a finite grid inertia system described by a cage model [21]-[27] as discussed in [12]. However, energy grid systems are also described via a Kuramoto-like model [12], [17]-[20]

$$\ddot{\theta}_i + \beta_i(\dot{\theta}_i - \Omega) + \xi_i\sin(\theta_i - \theta_j) = \bar{\tau}_i \ (i,j = grid,gen), (A7)$$

where $\beta_{grid} = K^K_{grid}/J_{grid}, \beta_{gen} = K^K_{gen}/J_{gen}$ are normalized damping parameters. Notice that here the damping parameters $K^K_{grid}, K^K_{gen}$ are in general not equal. If $K^K_{grid} = K^K_{gen} = K_D$, the Kuramoto-like model can also be analyzed using (A4) for $\delta(t) = \theta_{grid}(t) - \theta_{gen}(t)$. For infinite grid inertia, the cage and Kuramoto-like models both yields the same results.

### B. Calculations for Section 3

We describe the calculation of the stabilizing signal $V_{PSS}$, (7), and the output signal of the AVR $V_{out}$, (8), for the linear transient response of the generator. We consider a the generalized form of the input signal given by (1) so that this can be used to describe the rotor speed deviation, the frequency deviation of the bus voltage or the electrical power output [15]. The most common input to the PSS1A [15] is the rotor speed deviation. Then the calculations significantly simplify since this signal can be represented as a single decaying sine wave so that $b_0 = 0$, $V_\infty = 0$ (see (3)). This consideration also applies to the frequency deviation of the bus voltage (see (4)). However, for an input signal corresponding to the electrical power output we will invariably have $a_0 \neq 0$, $b_0 \neq 0$, $V_\infty = P_{max} \sin \delta_{II}$ (see (6)). When considering the rotor speed deviation $-\dot{\delta}(t)$ due to a disturbance occurring at $t = 0$, the deviation is zero before the event, and afterwards a function which again relaxes to zero. However, with the rotor angle $\delta(t)$ or the electrical power output $P_{el}(t) = P_{max} \sin \delta(t)$ following a disturbance, both $\delta(t)$ and $P_{el}(t)$ remain at a constant (typically nonzero) level up to that point (i.e., $\dot{\delta}(t \leq 0) = \dot{P}_{el}(t \leq 0) = 0$). Additionally, these signals will relax to another constant (typically nonzero) level. Therefore, the PSS1A will receive DC signals at both the intervals $t \leq 0$ and $t \to \infty$. Since the first block of the PSS1A acts as a *low pass filter*, these DC signals will be preserved so that

$$V_1(t \leq 0) = V_{in}(t \leq 0), \quad V_1(t \to \infty) = V_{in}(t \to \infty). \quad \text{(B1)}$$

In contrast, the washout filter (second block) acts as a *high pass filter* eliminating the DC signals so that

$$V_2(t \leq 0) = 0, \quad V_2(t \to \infty) = 0. \quad \text{(B2)}$$

Although the two lead-lag compensators (third and fourth blocks) preserve the DC signal, nevertheless due to the washout filter

$$V_{PSS}(t \leq 0) = V_3(t \leq 0) = V_2(t \leq 0) = 0, \quad \text{(B3)}$$

$$V_{PSS}(t \to \infty) = V_3(t \to \infty) = V_2(t \to \infty) = 0. \quad \text{(B4)}$$

In accordance with the first cascaded block of Fig. 2 (a) representing a first-order low-pass filter, the intermediate signal $V_1$ in the s-domain is

$$sT_6 \tilde{V}_1(s) + \tilde{V}_1(s) = \tilde{V}_{in}(s) \quad \text{(B5)}$$

or in the time domain using the inverse Laplace transform

$$T_6 \frac{dV_1(t)}{dt} + V_1(t) = V_{in}(t), \quad \text{(B6)}$$

where the Laplace transform is defined as

$$\tilde{V}(s) = \int_0^\infty V(t) e^{-st} dt. \quad \text{(B7)}$$

As (B6) is a first order linear differential equation,

$$V_1(t) = a_1 e^{-\lambda t} \sin \omega_0 t + b_1 e^{-\lambda t} \cos \omega_0 t + c_1 e^{-t/T_6} + V_\infty \quad \text{(B8)}$$

where the various coefficients are (noting (B1))

$$a_1 = \frac{(1 - \lambda T_6) a_0 + \omega_0 T_6 b_0}{(1 - \lambda T_6)^2 + (\omega_0 T_6)^2},$$

$$b_1 = \frac{-\omega_0 T_6 a_0 + (1 - \lambda T_6) b_0}{(1 - \lambda T_6)^2 + (\omega_0 T_6)^2},$$

$$c_1 = b_0 - b_1.$$

Likewise, we have the intermediate signals

$$V_2(t) = a_2 e^{-\lambda t} \sin \omega_0 t + b_2 e^{-\lambda t} \cos \omega_0 t + c_2 e^{-t/T_6} + d_2 e^{-t/T_5}, \quad \text{(B9)}$$

$$V_3(t) = a_3 e^{-\lambda t} \sin \omega_0 t + b_3 e^{-\lambda t} \cos \omega_0 t + c_3 e^{-t/T_6} + d_3 e^{-t/T_5} + e_3 e^{-t/T_2}, \quad \text{(B10)}$$

$$V_R(t) = a_R e^{-\lambda t} \sin \omega_0 t + b_R e^{-\lambda t} \cos \omega_0 t + c_R e^{-t/T_6} + d_R e^{-t/T_5} + e_R e^{-t/T_2} + f_R e^{-t/T_4} + s_R, \quad \text{(B11)}$$

as well as the stabilizing signal (7) and the output of the AVR (8), where the coefficients are for $V_2(t)$

$$a_2 = K_S T_5 \frac{(\omega_0^2 T_5 - \lambda(1 - \lambda T_5)) a_1 - \omega_0 b_1}{(1 - \lambda T_5)^2 + (\omega_0 T_5)^2},$$

$$b_2 = K_S T_5 \frac{\omega_0 a_1 + (\omega_0^2 T_5 - \lambda(1 - \lambda T_5)) b_1}{(1 - \lambda T_5)^2 + (\omega_0 T_5)^2},$$

$$c_2 = c_1 K_S T_5 / (T_5 - T_6),$$

$$d_2 = -b_2 - c_2,$$

for $V_3(t)$

$$a_3 = \frac{((1 - \lambda T_2)(1 - \lambda T_1) + \omega_0^2 T_2 T_1) a_2 + \omega_0 (T_2 - T_1) b_2}{(1 - \lambda T_2)^2 + (\omega_0 T_2)^2},$$

$$b_3 = \frac{\omega_0 (T_1 - T_2) a_2 + ((1 - \lambda T_2)(1 - \lambda T_1) + \omega_0^2 T_2 T_1) b_2}{(1 - \lambda T_2)^2 + (\omega_0 T_2)^2},$$

$$c_3 = c_2 (T_1 - T_6) / (T_2 - T_6),$$

$$d_3 = d_2 (T_1 - T_5) / (T_2 - T_5),$$

$$e_3 = -b_3 - c_3 - d_3,$$

for $V_{PSS}(t)$

$$a_4 = \frac{((1 - \lambda T_4)(1 - \lambda T_3) + \omega_0^2 T_4 T_3) a_3 + \omega_0 (T_4 - T_3) b_3}{(1 - \lambda T_4)^2 + (\omega_0 T_4)^2},$$

$$b_4 = \frac{\omega_0 (T_3 - T_4) a_3 + ((1 - \lambda T_4)(1 - \lambda T_3) + \omega_0^2 T_4 T_3) b_3}{(1 - \lambda T_4)^2 + (\omega_0 T_4)^2},$$

$$c_4 = c_3 (T_3 - T_6) / (T_4 - T_6),$$

$$d_4 = d_3 (T_3 - T_5) / (T_4 - T_5),$$

$$e_4 = e_3 (T_3 - T_2) / (T_4 - T_2),$$

$$f_4 = -b_4 - c_4 - d_4 - e_4,$$

for $V_R(t)$

$$a_R = \frac{(\omega_0^2 T_N - \lambda(1 - \lambda T_N)) a_4 + \omega_0 b_4}{T_N (\lambda^2 + \omega_0^2)},$$

$$b_R = \frac{-\omega_0 a_4 + (\omega_0^2 T_N - \lambda(1 - \lambda T_N)) b_4}{T_N (\lambda^2 + \omega_0^2)},$$

$$c_R = (1 - T_6 / T_N) c_4, \quad d_R = (1 - T_5 / T_N) d_4,$$

$$e_R = (1 - T_2 / T_N) e_4, \quad f_R = (1 - T_4 / T_N) f_4,$$

$$s_R = -b_R - c_R - d_R - e_R - f_R,$$

and for $V_{out}(t)$

$$a_{out} = \frac{(1 - \lambda T_S) a_R + \omega_0 T_S b_R}{(1 - \lambda T_S)^2 + (\omega_0 T_S)^2},$$

$$b_{out} = \frac{(1 - \lambda T_S) b_R - \omega_0 T_S a_R}{(1 - \lambda T_S)^2 + (\omega_0 T_S)^2},$$

$$c_{out} = c_R T_6 / (T_6 - T_S), \quad d_{out} = d_R T_5 / (T_5 - T_S),$$

$$e_{out} = e_R T_2 / (T_2 - T_S), \quad f_{out} = f_R T_4 / (T_4 - T_S),$$

$$g_{out} = -b_{out} - c_{out} - d_{out} - e_{out} - f_{out} - s_R.$$

## C. Calculations for Nonlinear Response

The calculations for the coefficients in (11) and (12) corresponding to the input signal (9) are the same as for the input signal (1), described in Appendix B. Since these calculations are easily reproduced, we do not give them. Here the signals $V_{PSS}(t)$ and $V_{out}(t)$ are given by (11) and (12), respectively, where the various constants are

$$a_{1j} = \frac{a_{0j}}{1+\lambda_j T_6}, \quad c_1 = \sum_j (a_{0j} - a_{1j}),$$

$$a_{2j} = \frac{K_S \lambda_j T_5}{1+\lambda_j T_5} a_{1j}, \quad c_2 = -\frac{K_S T_5}{T_6 - T_5} c_1,$$

$$d_2 = -\sum_j a_{2j} + c_2, \quad a_{3j} = \frac{1+\lambda_j T_1}{1+\lambda_j T_2} a_{2j},$$

$$c_3 = \frac{T_6 - T_1}{T_6 - T_2} c_2, \quad d_3 = \frac{T_5 - T_1}{T_5 - T_2} d_2,$$

$$e_3 = -\sum_j a_{3j} - c_3 - d_3,$$

$$a_{4j} = \frac{1+\lambda_j T_3}{1+\lambda_j T_4} a_{3j}, \quad c_4 = \frac{T_6 - T_3}{T_6 - T_4} c_3,$$

$$d_4 = \frac{T_5 - T_3}{T_5 - T_4} d_3, \quad e_4 = \frac{T_2 - T_3}{T_2 - T_4} e_3,$$

$$f_4 = -\sum_j a_{4j} - c_4 - d_4 - e_4,$$

$$a_{Rj} = \frac{1+\lambda_j T_N}{\lambda_j T_N} a_{4j}, \quad c_R = \frac{T_N - T_6}{T_N} c_4,$$

$$d_R = \frac{T_N - T_5}{T_N} d_4,$$

$$e_R = \frac{T_N - T_2}{T_N} e_4, \quad f_R = \frac{T_N - T_4}{T_N} f_4,$$

$$s_R = -\sum_j a_{Rj} - c_R - d_R - e_R - f_R,$$

$$a_{out,j} = \frac{1}{1+\lambda_j T_S} a_{Rj},$$

$$c_{out} = \frac{T_6 c_R}{T_6 - T_S}, \quad d_{out} = \frac{T_5 d_R}{T_5 - T_S},$$

$$e_{out} = \frac{T_2 e_R}{T_2 - T_S}, \quad f_{out} = \frac{T_4 f_R}{T_4 - T_S},$$

$$g_{out} = -\sum_j a_{out,j} - c_{out} - d_{out} - e_{out} - f_{out} - s_R.$$

ACKNOWLEDGMENT

We thank Dr W. J. Dowling for a critical reading of the paper.